\input epsf
\documentstyle[aps]{revtex}
\title{A dynamical study of fission process and estimation of prescission
neutron multiplicity}
\author{Asish~K.~Dhara, Kewal~Krishan, Chandana~Bhattacharya\cite{pa} and
Sailajananda~Bhattacharya }
\address{Variable Energy Cyclotron Centre, 1/AF Bidhan Nagar, Calcutta-
700 064, India}
\begin{document}
\maketitle
\begin{abstract}

The  dynamics  of  fission  has  been  studied  by  solving  Euler-Lagrange
equations with dissipation generated  through  one  and  two  body  nuclear
friction.   The   average   kinetic  energies  of  the  fission  fragments,
prescission neutron multiplicities and the mean energies of the prescission
neutrons have been calculated and compared  with  experimental  values  and
they agree quite well. A single value of friction coefficient has been used
to  reproduce  the  experimental  data  for  both  symmetric and asymmetric
splitting of the fissioning  systems  over  a  wide  range  of  masses  and
excitation  energies.  It  has  been  observed  that a stronger friction is
required in the saddle to scission region as compared to that in the ground
state to saddle region.

\end{abstract}

\pacs{PACS number(s): 25.70Jj,25.70Gh}

\section{Introduction}
\label{intro}

The  studies  of  fission dynamics has become a subject of current interest
due to availability of recent experimental data for prescission neutron
multiplicities in heavy ion induced fusion--fission reactions
\cite{e1,e2,e3,e4,e5,e6,e7,e8,e9,e10}. From the analysis of the data, it is
by and large established that standard statistical theory could not account
for  the  observed  large  multiplicity  of  the prescission neutrons. This
discrepency is, nowadays, believed to be of dynamical origin, and thus  led
to  enormous activities to understand theoretically the dynamics of fission
process \cite{abe1,abe2,mav,frob,pom,fp1,fp2,fp3,nix,feld,dha,cb}.

Dynamics  of  fission consists in the study of gradual change of shape of a
fissioning compound nucleus. The shape is globally characterised  in  terms
of  elongation  parameter, the neck radius and mass ratio of two fragments.
These variables  are  usually  referred  to  as  collective  variables  and
dynamics  is  understood  in  terms  of  the  evolution of these collective
variables. It has been observed that time taken  by  the  compound  nucleus
from  its state of formation to the state when it gets dissociated into the
two fragments is 'too short' when the shape evolves under the effect of the
conservative forces derived from the Coulomb  and  surface  energy  of  the
compound  nucleus.  With the average neutron evaporation rate determined by
excitation  of  the  compound  nucleus,  such  large  prescission   neutron
multiplicity data can not be explained in any manner with this 'short' time
scale  of  fission  process.  This  feature  leads  one to believe that the
dynamical process of shape evolution  gets  inhibited  for  a  considerable
amount of time. This retardation in the dynamics of collective variables is
effected  through  a  mechanism  of  dissipation by assuming the fissioning
nucleus  as  a  liquid  drop.  The  dissipation  is  usually  realised   by
introducing  a  friction term in the dynamics of shape evolution. Thus, the
explanation of the  observed  large  neutron  multiplicity  points  to  two
important  features  of  the fission process; the time scale of fission and
the viscosity of the  nuclear  fluid.  The  dynamics  of  fission  is  then
picturised  as a dissipative process where initial energy of the collective
variables get dissipated into the internal degrees of  freedom  of  nuclear
fluid  giving rise to the increase in internal excitation energy, which, in
turn, is responsible for the evaporation of prescission neutrons.

There have been different approaches to study this problem. One approach is
to   solve   the   Langevin   equations   for   the   collective  variables
\cite{abe1,abe2,frob,pom}. In this approach,  one  assumes  the  collective
variable  as  the 'Brownian particle' interacting stochastically with large
number of internal degrees of freedom constituting the surrounding  'bath'.
The  systematic  dissipative force is assumed to be derived from the random
force averaged over a time larger than the collisional time  scale  between
collective  and internal degrees of freedom. The random part is modelled as
a gaussian white  noise  which  causes  the  fluctuation  of  the  physical
observables  of the fission process such as kinetic energies, yields of the
fission  fragments  etc.  In   the   other   approach,   one   solves   the
multi-dimensional  Fokker-Planck  equation  \cite{fp1,fp2,fp3},  which is a
differential version of Langevin equation. In both of these approaches, the
calculated average kinetic energies, average yield of the fission fragments
and neutron multiplicity compare more or  less  well  with  the  respective
experimental data.

 Following  an  alternative  approach,  we  proposed  a  dynamical model of
fission \cite{dha,cb}, which could explain  fairly  well  various  features
observed  in fusion-fission reactions of lighter systems. In that model, we
solved the Euler-Lagrange equations for the collective variables instead of
solving the Fokker-Planck  or  Langevin  equations.  The  fluctuations  are
introduced  at  the initial level of the dynamics by random partitioning of
available energy of the 'nascent' compound nucleus between  the  collective
and  the  intrinsic  degrees  of  freedom and attributing the former to the
generator of the displacement of the  collective  variable  as  an  initial
condition  of Euler-Lagrange equation. The random initial momentum given to
the collective variable gives rise to different trajectories. Some of these
trajectories would cross the fission barrier and have the  'fission  fate'.
The main difference between our approach and those of the others is that in
our  approach  the  dynamics is deterministic with randomisation at initial
level, while in other approaches, the  dynamics  is  mainly  stochastic  in
nature.  However,  in all these approaches the dissipation is introduced in
terms of frictional forces.

In  our  earlier calculations \cite{dha,cb} we assumed a schematic shape of
the fissioning system which comprised of two leptodermous spheres connected
by a cylindrical neck.  This  particular  idealisation  was  introduced  by
Swiatecki  \cite{swi}  to  simplify  the calculation of various ingredients
such as conservative forces etc. After the  saddle  point,  the  neck  gets
constricted  and at the scission point the spheres get detached forming two
fission fragments. The calculation with such simple shape produced  results
which  agree  quite  well  with the experimental data, particularly for the
nuclei lying below the Bussinaro-Gallone point. It is, however, known  that
for heavy nuclei lying above the Businaro-Gallone point, the fission shapes
are  highly  deformed,  and  elongation  is  quite large before the nucleus
reaches  the  scission  point.  Obviously,  the  simple   schematic   shape
parametrisation \cite{swi} used earlier would not be expected to mimic such
large deformed shapes. Besides, the dissipation of collective energy to the
internal  nucleonic  degrees  of freedom could happen through two different
mechanisms. One is due to two-body collision of  the  nucleons  inside  the
nuclear  fluid,  the  other  is  due to collision of nucleons with changing
surface of the nuclear fluid, which is more commonly known as  one-body  or
'Wall'  friction  and  both  of them depend very sensitively on the surface
profile of the fissioning system. Hence, with these considerations in mind,
in the present paper, we have used a more realistic  shape  parametrisation
of  the heavy fissioning systems to study the temporal evolution of fission
shapes, and, subsequently, to calculate various observables of the  fission
process, such as prescission neutron multiplicities, total kinetic energies
etc.

The  present  paper  has  been  arranged as follows. In Sec.~\ref{model} we
describe the model and the  statistics  used  in  calculating  the  average
kinetic  energies,  prescission  neutron multiplicities etc. The results of
the calculation are discussed in  Sec.~\ref{results}.  Finally,  concluding
remarks are given in Sec.~\ref{conclud}.

\section{the model}
\label{model}

The  present  model is a generalisation of the schematic model developed by
us \cite{dha} to study the fusion-fission process for light nuclear systems
lying below the Businaro-Gallone point. The schematic  parametrisation  for
the  evolution of fission shapes used in the earlier work has been replaced
by a more realistic parametrisation for the same  that  is  applicable  for
heavier nuclei with fissility parameters above the Businaro-Gallone value.

\subsection{The shape}
The shape of the nuclear surface is assumed to be of the form

\begin{mathletters}
\begin{equation}
\label{eq.1}
\rho^2(z) = c^{-2}(c^2-z^2)(A+Bz^2+\alpha z c),
\end{equation}

where the coefficients $A$ and $B$ are defined as

\begin{equation}
\label{eq.2}
A=c^{-1}-B/5,
\end{equation}

\begin{equation}
\label{eq.3}
B=(c-1)/2.
\end{equation}
\end{mathletters}

This  is  a  specific  form  of  the  surface  introduced  by Brack et. al.
\cite{bra}. The quantity $c$ corresponds to the elongation and the quantity
$\alpha$ is a  parameter  which  depends  upon  the  asymmetry  ($a_{sym}$)
defined below. We may note that the surface cuts the $z$-axis at $z=\pm c$,
so  that  the  surface  to  surface separation along the elongation axis is
$2c$. There are three  real  points  at  which  the  derivative  of  $\rho$
vanishes; two of them correspond to maximum of $\rho$ and between these two
maxima  one  minimum occurs at $z=z_{min}$. The portion of volume contained
in $z=-c$ to $z=z_{min}$ is defined as left lobe and remaining  portion  of
the  volume  is  referred  to  as right lobe. Asymmetry ($a_{asy}$) is then
defined as

\begin{equation}
\label{eq.6}
a_{asy}=\frac{(A_R  - A_L)}{A_{CN}}
\end{equation}

where $A_{CN}$ is the compound nucleus mass, and $A_R, \ A_L$ correspond to
the  masses  of  the  right  and  left  lobes,  respectively. The parameter
$\alpha$ is related  to  the  asymmetry  $a_{asy}$  through  the  following
relation

\begin{equation}
\label{eq.7}
\alpha = .11937 a_{asy}^2 + .24720 a_{asy}.
\end{equation}

As  the shape changes gradually, the coordinates of the two maxima and that
of the minimum change. The scission point is defined when the minimum point
touches the $z$-axis and it is given by

\begin{equation}
\label{eq.4}
A  -\frac{c^2\alpha^2}{4  B}  =  0 .
\end{equation}

Therefore,  the  value  of $c$ at which scission occurs depends on $\alpha$
and the dependence is given by

\begin{equation}
\label{eq.5}
c_{sc} = -2.0 \alpha^2  + .032 \alpha  + 2.0917.
\end{equation}

\subsection{The dynamics}
\label{dyn}

The   dynamics   is  studied  by  calculating  the  semi-classical  fission
trajectories. The trajectories are obtained by solving the Euler-Lagrangian
equation \cite{dha},

\begin{mathletters}
\begin{eqnarray}
\label{eq.8}
\mu \ddot{r} -\frac{L^2}{\mu r^3} &=&-\gamma_r \dot{r} -\frac{\partial
(V_C+V_N)}{\partial r},\\
I_1 \ddot{\theta_1}&=&\gamma_t[g_2(\dot{\theta_2} - \dot{\theta}) +
g_1(\dot{\theta_1} - \dot{\theta})] g_1,\\
I_2 \ddot{\theta_2}&=&\gamma_t[g_2(\dot{\theta_2} - \dot{\theta}) +
g_1(\dot{\theta_1} - \dot{\theta})] g_2,\\
\dot{L}&=&-(I_1 \ddot{\theta_1} + I_2 \ddot{\theta_2}).
\end{eqnarray}
\end{mathletters}

The  quantities  $V_C$, $V_N$ represent the Coulomb and nuclear interaction
potentials  and  $\gamma_r$,  $\gamma_t$  are  the  radial  and  tangential
components of friction, respectively. $I_1, I_2$ are the moments of inertia
of the two lobes and $L$ refers to the relative angular momentum. $g_1$ and
$g_2$  are  the  distances of the centres of mass of the two lobes from the
centre  of  mass  of  the  composite  dinuclear   system   and   the   term
$[g_2(\dot{\theta_2} - \dot{\theta}) + g_1(\dot{\theta_1} - \dot{\theta})]$
represents  the  relative  tangential velocity of the two lobes \cite{dha}.
The variable $r$ is defined as the centre to centre  distance  between  the
two  lobes.  From  the generalised shape given by Eqn.~\ref{eq.1}, we first
construct the centres of mass of left and right lobes, and call them  $z_l$
and $z_r$ respectively. Then $r$ is defined as

\begin{equation}
\label{eq.9}
r =|z_l-z_r|  .
\end{equation}

The reduced mass parameter $\mu$, is obtained from the calculated masses of
the two lobes.

For the non-conservative part of the interaction, we would consider viscous
drag  arising  not  only  due  to  two  body  collision but also due to the
collisions of the nucleons with the wall or surface of the  nucleus.  Hence
$\gamma_r$  in  Eqn.~\ref{eq.8}  contains  two  parts;  $\gamma_r^{TB}$ and
$\gamma_r^{OB}$,  for  two-body  and   one-body   dissipative   mechanisms,
respectively.  Assuming the nucleus as an incompressible viscous fluid, and
for nearly irrotational hydrodynamical flow, $\gamma_r^{TB}$ is  calculated
by use of the Werner-Wheeler method \cite{nix,feld} and is given by

\begin{mathletters}
\begin{equation}
\label{eq.10}
\gamma_r^{TB} = \pi \mu_0 \ R_{CN} f(\partial c/\partial x)
\int^{+c}_{-c} dz\rho^2[3A^{' 2}_c +\frac{1}{8}\rho^2A^{''2}_c]
\end{equation}

where

\begin{equation}
\label{eq.11}
A_c(z) = -\frac{1}{\rho^2(z)} \frac{\partial}{\partial c}
\int^z_{-c} dz'\rho^2(z').
\end{equation}
\end{mathletters}

The  quantities  $A'_c,A''_c$  are  the  first  and  second  derivatives of
$A_c(z)$  with  respect  to  $z$.  $\mu_0$  is  the  two   body   viscosity
coefficient. The factor $f(\frac{\partial c}{\partial x})$ is taken to be

\begin{equation}
\label{eq.12}
f(\frac{\partial c}{\partial x}) = (\frac{\partial c}{\partial x})^2 +
2(\frac{\partial c}{\partial x }),
\end{equation}

where $x=r/R_{CN}$, $R_{CN}$ being the radius of the compound nucleus. This
factor  is  a  consequence  of  the rotational symmetry of the shape ( \ref
{eq.1}) around the elongation axis. The variable $r$ is already defined and
the relationship between $c$ and $x$ is found to be

\begin{mathletters}
\begin{equation}
\label{eq.13}
c = p x^2 + q x +\tilde{r}(\alpha),
\end{equation}

where, $p = -.15901$, $q = 1.03749$, and  $\tilde{r}$ is given by,

\begin{equation}
\label{eq.14}
\tilde{r} = -1.228 \alpha^2 - .01896 \alpha + .45956 .
\end{equation}
\end{mathletters}

 The  tangential friction $\gamma_t^{TB}$ is calculated using the following
relation \cite{dha},

\begin{equation}
\label{g_tb}
\gamma_t^{TB} = (\frac{\partial c}{\partial n})^2 \gamma_r^{TB}.
\end{equation}

\begin{figure}[h]
\centering
\epsfysize=8 cm
\epsffile[65 226 536 441]{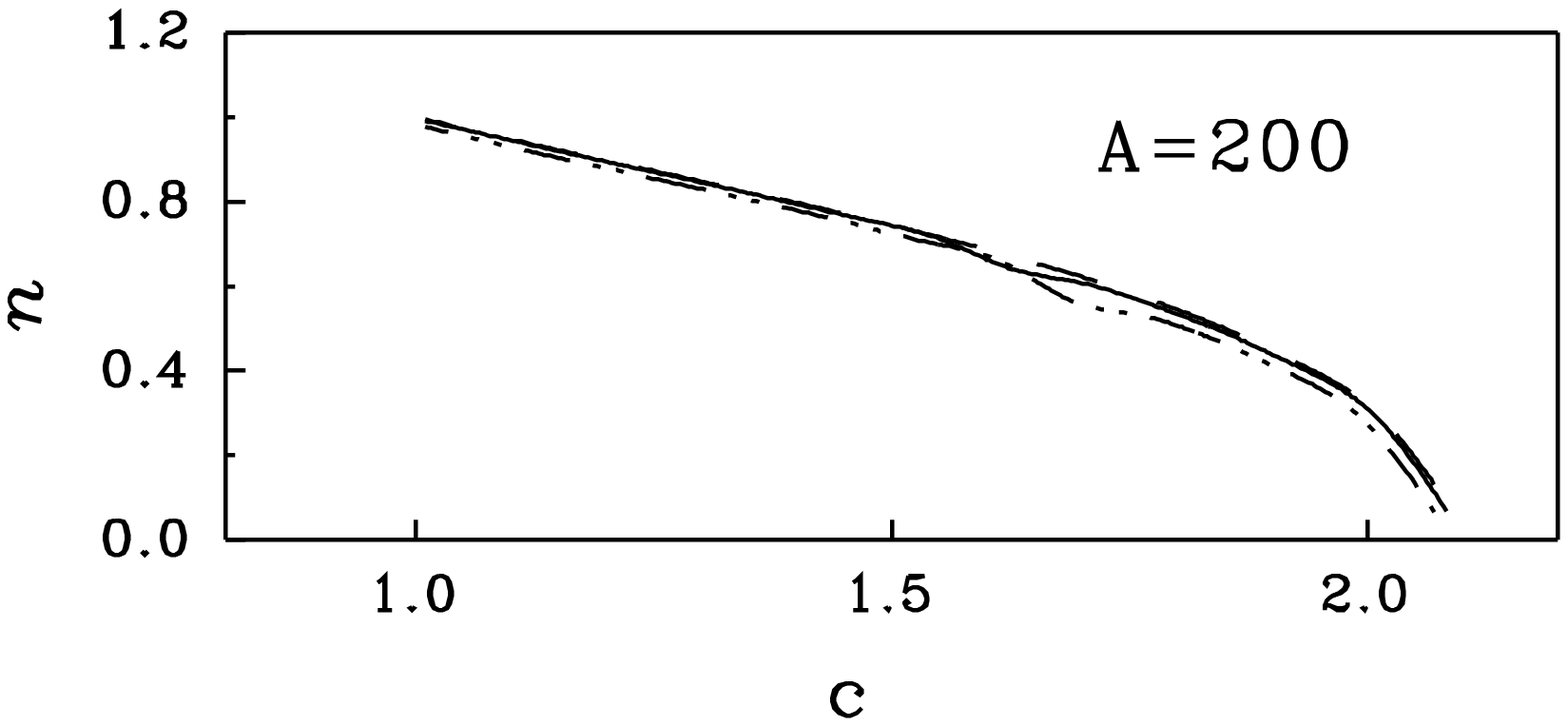}
\caption{Variation  of neck radius $n$ with $c$ for different values of the
parameter $\alpha$. Dash, solid and dash-dot lines correspond to  $\alpha$=
0, 0.05 and 0.10, respectively.}

\label{fig_neck}
\end{figure}

The  quantity $n$ refers to the neck radius of the composite shape given by
Eqn.~\ref{eq.1}. It is defined as the value of $\rho$ where $\rho^2$ has  a
minima. The variation of $n$ with $c$ for different values of the parameter
$\alpha$  is  shown  in  Fig.~\ref{fig_neck}. It is evident from the figure
that for all values of $c$, the corresponding  values  of  $n$  are  nearly
independent  of  $\alpha$,  and  $n$  is  found to be related to $c$ by the
following relation,

\begin{equation}
\label{eq_neck}
n = -1.047 c^3 + 4.297 c^2 - 6.309 c + 4.073
\end{equation}

One  body dissipative force, $F_{dis}$, is obtained from the rate of energy
dissipation, $E_{dis}$, by

\begin{equation}
\label{eq.15}
F_{dis} = -\frac{\partial}{\partial \dot{x}} E_{dis}(x)
\end{equation}

where  $\dot{x}$  refers  to the rate of change of $x$ with respect to time
and $E_{dis}(x)$ is the rate of energy dissipation at $x$ given by

\begin{equation}
\label{eq.16}
E_{dis} = \frac{1}{2} \rho_m \bar{v} \oint dS \ {\dot{\vec{e}_n}}^2,
\end{equation}

where  $\vec{e}_n$  is  the  unit  normal  direction  at  the  surface. The
integration is done over the whole surface. $\rho_m$ is the nuclear density
and $\bar{v}$ is the average nucleonic speed obtained from the formula

\begin{equation}
\label{eq.17}
\bar{v} = \surd(\frac{8k}{m\pi}) (E_{av}/a)^{1/4}
\end{equation}

with $E_{av}$ is the available energy and the level density parameter, $a$,
is  taken  to  be  $A_{CN}/10$.  For  the  generalised  shape (\ref{eq.1}),
one-body friction, $\gamma_r^{OB}$, is obtained as

\begin{equation}
\label{eq.18}
\gamma_r^{OB} = 2\pi \ \rho_m \bar{v} R_{CN}^2  f(\partial c/\partial x)
\int^{+c}_{-c} dz \ \rho \ [1+ \ \rho\prime^2]^{-1/2}
[A_c\rho\prime + (1/2) \ \rho A'_c]^2
\end{equation}

where  $\rho\prime, \ A'_c$ are the derivatives of $\rho ,A_c$ with respect
to $z$ and all other quantities are defined earlier. The tangential friction
 $\gamma_t^{OB}$ is calculated in a similar way as in Eqn.~\ref{g_tb}.

After  the  formation,  the  compound  nucleus  is  in  the  minimum of the
potential energy surface and it is assumed that the  total  initial  energy
available  in  the fusion process is equilibrated to allow the system to be
described in terms of a thermodynamic state. Some part of it may be  locked
in  rotational energy of the compound nucleus and the remaining part is the
available excitation  energy.  However,  what  fraction  of  the  available
excitation  energy will be converted into the collective excitation leading
to the temporal evolution of fission shapes, is not known apriori.  In  our
model,  it  is  assumed  that a random fraction of this available energy is
imparted to the collective degrees of freedom which initiates the  dynamics
of the fission process. The initial conditions of $r$ and $\dot{r}$ are

\begin{mathletters}
\begin{equation}
\label{eq.19}
r(t=0) = r_{min},
\end{equation}

\begin{equation}
\label{eq.20}
\dot{r}(t=0) = (E^* R_N/2\mu)^{1/2},
\end{equation}
\end{mathletters}

where  $R_N$  is  a  random number between 0 and 1 from uniform probability
distribution. The available excitation energy, $E^*$, is given by

\begin{equation}
\label{eq.21}
E^* = E_{cm} + Q_{ent} -L^2/2I_{CN},
\end{equation}

where $E_{cm}$ and $Q_{ent}$ are the centre of mass energy and $Q$-value in
the  entrance  channel,  and,  $I_{CN}$  is  the  moment  of inertia of the
compound nucleus. The initial conditions of the  angle  variables  at  time
$t=0$ are

\begin{mathletters}
\begin{equation}
\theta(0)    =    \theta_1(0)   =   \theta_2(0)   =   \dot{\theta_1}(0)   =
\dot{\theta_2}(0)
= 0,
\end{equation}

\begin{equation}
\dot{\theta}(0) = L/I_{CN}
\end{equation}
\end{mathletters}

 The  dynamical  evolution  starts at $r=r_{min}$, where the minimum of the
potential energy occurs. Once it reaches the saddle point or the top of the
barrier, it is almost certain that it will reach the scission point.

\subsection{The neutron multiplicity}
\label{sec_nmul}

The  emission  of  the  prescission neutrons is incorporated in the present
model as follows. During the temporal evolution of the  fission  trajectory
the  intrinsic  excitation  of  the system is calculated at each time step.
Correspondingly, the neutron decay width  at  that  instant,$\Gamma_n$,  is
calculated  using the relation $\Gamma_n = \hbar W_n$, where the decay rate
$W_n$ is given by

\begin{equation}
\label{eq.22}
W_n = \int^{E_{max}}_0 dE \frac{d^2\Pi_n}{dEdt}.
\end{equation}

The rate of decay $A\rightarrow A-1+n$ in an energy interval $[E,E+dE]$ and
a time interval $[t,t+dt]$, $\frac{d^2\Pi_n}{dEdt}$, is given by

\begin{equation}
\label{eq.23}
\frac{d^2\Pi_n}{dEdt} = (1/\pi^2\hbar^3) E \sigma_{inv}\mu_r \ \frac
{\omega_{A-1}(E^*_{A-1})}{\omega_A(E^*_A)}.
\end{equation}

The  quantity  $\omega_A(E^*)$  is  the  level density for the nucleus with
atomic number $A$  and  excitation  energy  $E^*$.  $\sigma_{inv}$  is  the
inverse  cross  section for the reaction $(A-1)+n\rightarrow A$ and $\mu_r$
is the reduced mass of $(A-1,n)$ system. The upper limit of integration  in
Eqn.~\ref{eq.22} is given by

\begin{equation}
\label{eq.24}
E_{max} = E^* + B_A -(B_{A-1} + B_n),
\end{equation}

where $B_A$ is the binding energy of the nucleus with atomic number $A$ and
$B_n$  is  the  neutron  separation  energy. The emission of neutron in the
evolution of trajectory is now conceived by fixing a criterion. We evaluate
the ratio of neutron decay time $\tau_n(=\hbar/\Gamma_n)$ and the time step
$\tau$ of the calculation. The  ratio  $\tau/\tau_n$  is  compared  with  a
random  number $R_N$ from a uniform probability distribution. The criterion
of emission of a neutron at  random  time  is  fixed  with  the  rule  that
whenever

\begin{equation}
\label{eq.25}
\tau/\tau_n > R_N
\end{equation}

the  emission  of  a neutron takes place. If condition (\ref{eq.25}) is not
satisfied, no emission of neutron takes place. The probability of  emission
of  a  neutron  in  time $\tau$ is $(\tau/\tau_n)$. The time step $\tau$ is
chosen in such a way that it satisfies the condition $\tau/\tau_n  \ll  1$.
This  suggests  that  the  evaporation  is  a  Poisson  process  leading to
exponential decay law with half life $\tau_n$ \cite{cb}.  Consequently  the
probability  of  emission  of  two or more neutrons in time $\tau$ would be
extremely small.

The  kinetic  energy  of  the  emitted  neutron is extracted through random
sampling technique. For this purpose, it is assumed that the system  is  in
thermal  equilibrium at each instant of time $t$, and therefore, the energy
distribution of the emitted neutrons may be  represented  by  a  normalised
Boltzmann  distribution  corresponding  to the instantaneous temperature of
the system. From a uniformly distributed random number  sequence  \{$x_n$\}
in  the  interval  [0,1],  we  construct  another  random  number  sequence
\{$y_n$\}  with  probability  distribution   $f(y)$,   where   $f(y)   \sim
\exp(-\beta(t)  y)$ is a normalised Boltzmann distribution corresponding to
the temperature $\beta(t)$ at any instant of time $t$. Then,  the  sequence
\{$y_n$\} is obtained from the sequence \{$x_n$\} by the relation,

\begin{mathletters}
\begin{equation}
y(x) = F^{-1}(x),
\end{equation}

where, $F^{-1}$ is the inverse of the function $F(y)$, which is given by,

\begin{equation}
x = \int_0^y f(y) dy = F(y).
\end{equation}
\end{mathletters}

The inverse function is computed numerically by forming a table of integral
values.  The energy of the emitted neutron is given by $E_n = y E_n^{max}$,
where $E_n^{max}$ is chosen in such a way that the Boltzmann probability at
that energy is negligible for all instants of time $t$. After the  emission
of  the  neutron,  the  intrinsic excitation energy is recalculated and the
trajectory is continued. In this way for each angular momentum $l$  of  the
compound  system,  the average number of emitted neutrons per fission event
$<M_n>_l$, is calculated.

\subsection{The statistics}

As argued previously, in the present model randomness is introduced only at
the  initial  level  when  the  compound  nucleus  is at the minimum of the
potential energy surface. A random fraction of excitation energy  is  given
to  the  collective  degrees of freedom. As a consequence, all trajectories
would not be able to cross the barrier and would  not  have  fission  fate.
Apart  from  that,  there  is a parameter $\alpha$ in Eqn.~\ref{eq.1} which
decides  the  final  asymmetry  of  fission  fragments   uniquely   through
Eqn.~\ref{eq.7}.  For obtaining different asymmetry, one should introduce a
probability distribution $P_l(\alpha)$ at the initial level. It is  natural
to  assume  $P_l(\alpha)$  to  be  proportional  to  the  density of states
available for that $\alpha$. Thus $P_l(\alpha)$ is taken to be

\begin{mathletters}
\begin{equation}
\label{eq.26}
P_l(\alpha) \propto  u^{-2} exp[2(au)^{1/2}]
\end{equation}

where $a$ is the level density parameter and $u$ is given by

\begin{equation}
\label{eq.27}
u = E^* - V_{min}(\alpha)
\end{equation}
\end{mathletters}

and  $V_{min}(\alpha)$ is the minimum of the potential energy surface for a
given $\alpha$.

The  compound  nucleus  is  formed  from  the fusion process with different
angular momentum. The  dynamics  to  follow  after  its  formation  depends
intricately  on this angular momentum as seen from (\ref{eq.21}). Hence the
probability to cross the barrier would depend upon this  angular  momentum.
We  call  this probability $P_l(f,\alpha|l)$. This is obtained as the ratio
of number of trajectories crossing the barrier for given $\alpha$  and  $l$
and  the  total  number  of  trajectories  chosen. Therefore average of any
observable quantity, $O$, is given by

\begin{equation}
\label{eq.28}
<O> = \frac{\sum_{l=0}^{l=l_{cr}} (2l+1)O(\alpha,l) P_l(f,\alpha|l)
P_l(\alpha)}{\sum_{l=0}^{l=l_{cr}} (2l+1) P_l(f,\alpha|l) P_l(\alpha)},
\end{equation}

where  the quantity $O$ may be any of the relevent observables of interest,
{\it e.g.}, kinetic energy, neutron multiplicity etc., and, $l_{cr}$ is the
critical angular momentum for fusion.

\section{Results and Discussions.}
\label{results}

A  large  amount of prescission neutron multiplicity data over a wide range
of excitation energies and masses of  the  compound  nucleus  is  presently
available in the literature. We have chosen a few representative systems
in the ranges of masses A$_{CN}$ $\sim$ 150--250, and, excitation energies
$E^*_{CN}$ $\sim$ 60--160 MeV. All the systems  considered  here
are   above  the  Businaro-Gallone  point  and  symmetric  fission  is  the
predominant mode of  decay.  Therefore,  the  theoretical  predictions  for
various  physical observables, {\it ie,} prescission neutron multiplicities
($n_{pre}$),  total  kinetic  energies  (TKE),  average  energies  of   the
prescission  neutrons ($<E_n>$), etc., have been made for the the symmetric
fission and confronted with the respective data. In addition,  some  recent
experiments  have been reported where fragment mass asymmetry dependence of
the related physical observables have been studied for a few of the systems
mentioned above. Explanation of such exclusive data is a crucial  test  for
any  theoretical  model  and it has not been, to the best of our knowledge,
attempted so far. Therefore, we have made  calculations  for  the  fragment
mass  asymmetry  dependence  of  some  of  the physical observables and the
results have been compared with the corresponding experimental data.

\subsection{Fission shapes and Friction form factors}

\begin{figure}[h]
\centering
\epsfysize=10 cm
\epsffile[32 181 470 710]{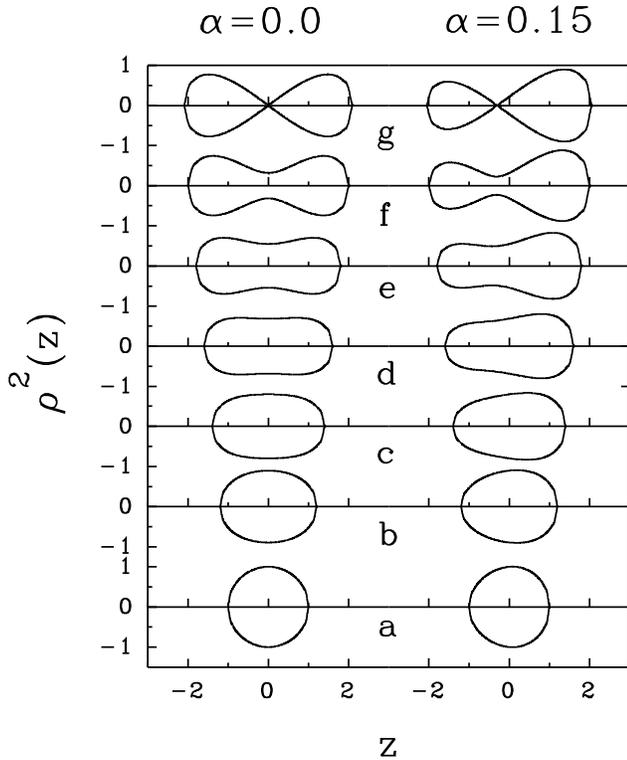}

\caption{Evolution  of  fission  shapes  for  a  compound  system  of  mass
$A_{CN}=200$ for $\alpha$ = 0.0 and 0.15. Figs. (a) to  (g)  represent  the
shapes  of  the  fissioning shapes for $c$ = 1, 1.2, 1.4, 1.6, 1.8, 2.0 and
$c_{sc}$, respectively.  'Negative'  values  of  $\rho^2$  are  the  mirror
reflection of the upper half about the elongation axis.}

\label{f_shape}
\end{figure}

The evolution of fission shapes for the symmetric ($\alpha=0.0$) as well as
asymmetric ($\alpha=0.15$) splitting of a representative compound system of
mass     A$_{CN}$=200     are     illustrated     in    Fig.~\ref{f_shape}.
Fig.~\ref{f_shape}a to Fig.~\ref{f_shape}g represent the gradual  evolution
of  the shape from the spherical ground state ($c=1$) to the scission point
($c=c_{sc}$). It may be pointed out that the centre to centre separation at
the scission point is $\sim 18$ fm which is much larger than what one  gets
in the schematic shape of \cite{swi}.

\begin{figure}[h]
\centering
\epsfysize=8 cm
\epsffile[56 207 438 590]{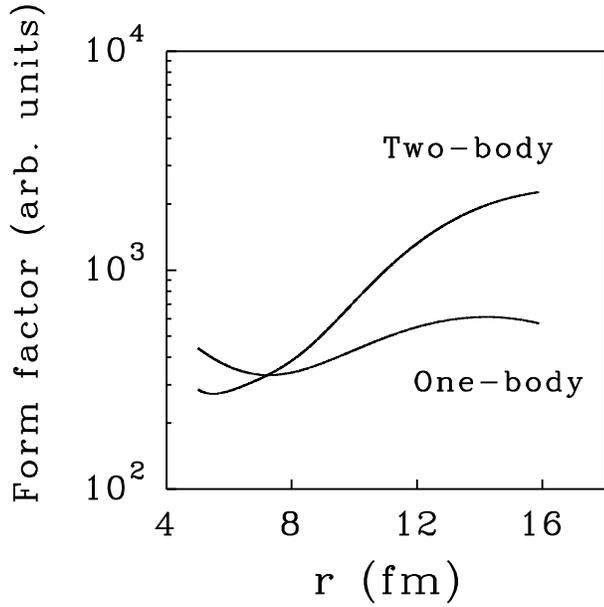}

\caption{Friction  form factors for one-body and two-body frictions plotted
as a function of the centre  to  centre  separation  $r$  between  the  two
symmetric fragments for a compound system of mass $A_{CN}=200$.}

\label{f_form}

\end{figure}

 In  the  present calculations, a combination of both one-body and two-body
frictions has been used to  calculate  the  fission  trajectories  and  the
fission  observables. In Fig.~\ref{f_form} form factors of the one-body and
two-body frictions are displayed as a function  of  the  centre  to  centre
separation between the two symmetric fragments for a typical system of mass
$A_{CN}=200$. It is seen from Fig.~\ref{f_form} that at smaller separations
(  when  the  shape  is  nearly mononuclear), one-body friction is stronger
whereas  at  larger  separations,  two-body  friction  dominates.  However,
one-body  friction  does  not  change  much with the increase in separation
between the fragments.

\subsection{Prescission Neutron Multiplicities}
\label{sec_mul}

As  prescission  neutron multiplicities depend on time scale of the fission
process  and  {\it  vis-a-vis}  on  the  magnitude  of  nuclear   friction,
theoretical  estimates  of  the  friction  coefficients are usually made by
reproducing  the  prescission  neutron  multiplicity  data  using  friction
coefficient   as   an   input  parameter  in  the  model.  In  the  present
calculations, one-body 'wall' friction has been used in the ground state to
saddle region, where nuclear shapes are nearly mononuclear. The strength of
the one-body friction used was attenuated to 10\% of  the  original  'wall'
value. This weakening of the wall friction has also been confirmed from the
study  of the role of chaos in dissipative nuclear dynamics \cite{spal}. In
the saddle to scission region, on the other hand, the  nuclear  dissipation
was  taken to be of two-body origin. For the two-body friction, the viscous
drag  was   calculated   in   the   framework   of   Werner-Wheeler   using
Eqn.~(\ref{eq.10})  and the value of the viscosity coefficient $\mu_0$ used
in the present calculation was (4 $\times10^  {-23}  MeV  \cdot  sec  \cdot
fm^{-3}$).   This   value   of   $\mu_0$   corresponds  to  0.06  TP  (  $1
TP=6.24\times10^{-23}MeV\cdot sec \cdot fm^{-3}$).

\begin{figure}[h]
\centering
\epsfysize=8 cm
\epsffile[76 110 431 553]{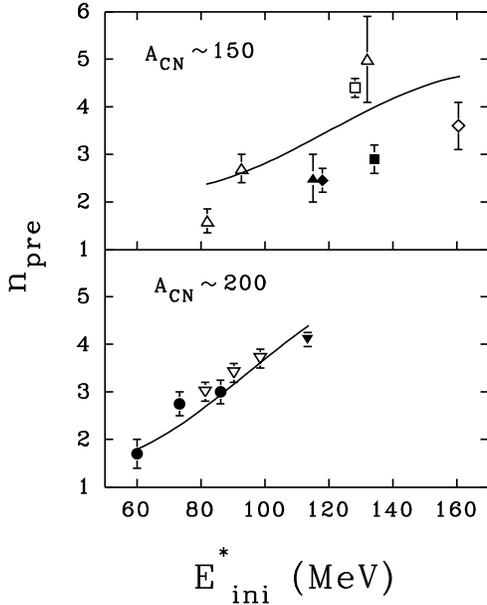}

\caption{Prescission  neutron  multiplicities  plotted as a function of the
initial excitation energy $E^*_{ini}$ of the compound nuclei of  masses
A$_{CN}  \sim$
150 ({\it upper half}), and A$_{CN} \sim$ 200 ({\it lower half}). The solid
curve is the present calculation. Different symbols correspond to different
sets   of   experimental  data,  ({\it  ie,}  filled  circle  $\rightarrow$
$^{28}$Si+$^{170}$Er    \protect\cite{e4},    open    inverted     triangle
$\rightarrow$   $^{19}$F+$^{181}$Ta   \protect\cite{e4},   filled  inverted
triangle $\rightarrow$ $^{18}$O+$^{197}$Au \protect\cite{e3}, open triangle
$\rightarrow$  $^{18}$O+$^{150}$Sm   \protect\cite{e4},   filled   triangle
$\rightarrow$    $^{24}$Mg+$^{134}$Ba   \protect\cite{e6},   open   diamond
$\rightarrow$   $^{16}$O+$^{142}$Nd   \protect\cite{e6},   filled   diamond
$\rightarrow$    $^{18}$O+$^{144}$Sm    \protect\cite{e3},    open   square
$\rightarrow$   $^{18}$O+$^{154}$Sm   \protect\cite{e3},   filled    square
$\rightarrow$ $^{18}$O+$^{124}$Sn \protect\cite{e3}).}

\label{fig_mul}
\end{figure}

The  calculated  prescission  neutron multiplicities have been displayed in
Fig.~\ref{fig_mul} as a function of the initial excitation  energy  of  the
compound nucleus for two different mass regions, {\it ie,} for $A_{CN} \sim
150$  ({\it  upper  half}),  and  $A_{CN} \sim 200$ ({\it lower half}). The
solid  curves  represent  the  results  of  the  present  calculations  and
different  symbols  correspond  to  different  sets  of  experimental  data
\cite{e3,e4,e6}. It is seen that for heavier systems ($A_{CN}  \sim  200$),
the  theoretical  predictions  are in good agreement with the corresponding
experimental  data.  For  lighter  systems   ($A_{CN}   \sim   150$),   the
experimental  points  are  somewhat  scattered  and  the  theory is seen to
reproduce quite well the average trend of the data. Here,  the  prescission
neutron  multiplicities  are  less and the uncertainties are more which are
reflected in the  larger  error  bars  of  the  experimental  measurements.
Moreover,  as  different  experimental  points belong to different compound
nuclei (different  symbols  in  the  figure),  additional  fluctuations  in
neutron  emission  due  to specific structure effects may not be ruled out.
For example, compound system $^{162}Yb$, formed in the reaction $^{18}$O  +
$^{144}$Sm  ({\it  filled diamond}), is quite neutron deficient compared to
$^{168}Yb$, formed  in  the  reaction  $^{18}$O  +  $^{150}$Sm  ({\it  open
triangle}).  Therefore,  neutron emission from the former is expected to be
somewhat less. In fact, such system dependent fluctuations of  the  average
neutron  multiplicities have also been observed when calculations were done
for specific systems (see text below). At very  high  excitation  energies,
the  observed  multiplicity ({\it open diamond}) was found to be lower than
the average theoretical trend. In this case, the incident energy was  quite
high ($>$ 10 MeV/nucleon), and the onset of preequilibrium emission process
may  not  be  ruled  out.  As  preequilibrium particles carry away a larger
amount of energy compared to the evaporated particles, the fused  composite
cools  down  faster  and subsequently leads to fewer emission of evaporated
particles \cite{e3}. As preeqilibrium emission has not been  considered  in
the  present  calculation,  the  model  predictions  of  multiplicities are
expected to be somewhat higher than the experimental  measurements  of  the
same at higher bombarding energies.

In Fig.~\ref{fig_ke}(a), we have plotted prescission neutron multiplicities
$n_{pre}$  as  function of the compound nuclear mass $A_{CN}$ for 158.8 MeV
$^{18}O$ induced reactions on different targets. The solid curve represents
the theoretical predictions whereas the filled circles  correspond  to  the
experimental  data  \cite{e3}.  It is seen that the observed multiplicities
increase  with  the  increase  in  $A_{CN}$  in  general  and   show   some
fluctuations in the vicinity of $A_{CN} \sim 160 - 170$, in particular. The
present  calculations  are  found to be quite successful in reproducing the
general trend of the data over the  whole  range  of  masses  studied.  The
fluctuations  in  the observed multiplicities, which may be due to specific
structure  effects,  as  discussed  earlier,  have  also  been   reproduced
qualitatively in the present model.

It  is  interesting  to  note that the present model with a single value of
friction coefficient, $\mu_0$ has been able to explain the general trend of
the data over the whole range of masses  and  excitation  energies  of  the
compound  nuclei  studied.  This  is  similar  to  what  one expects in the
Werner-Wheeler prescription,  which  predicts  that  the  reduced  friction
coefficient,  $\beta$  should be a universal function depending only on the
collective degrees  of  freedom  \cite{gon92}.  Earlier  attempts,  on  the
contrary,  have  shown  that  the  values  of the friction coefficients are
system dependent and  vary  over  a  wide  range  (typically,  the  reduced
friction coefficient, $\beta$, may have values between $2 \times 10^{-21} 
sec^{-1}$  and  $20  \times  10^{-21}  \  sec^{-1}$  \cite{mav}).  However,
Frobrich {\it et. al.} attempted to arrive at such a universal value of the
$\beta$ in the framework of modified Langevin equation model \cite{frob} by
using the  level  density  formula  proposed  by  Ignatyuk  {\it  et.  al.}
\cite{igna}.  They  had  to  use  a  constant value of $\beta$ (= $2 \times
10^{-21} \ sec^{-1}$) for the ground state to saddle followed by a  set  of
$\beta$ values proportional to the elongation parameter, which reaches upto
a  value  of $30 \times 10^{-21} \ sec^{-1}$ at the scission point in order
to reproduce the experimental data.

\subsection{Energy of emitted neutrons and TKE}
\label{sec_tke}

\begin{figure}[h]
\centering
\epsfysize=8 cm
\epsffile[75 206 512 655]{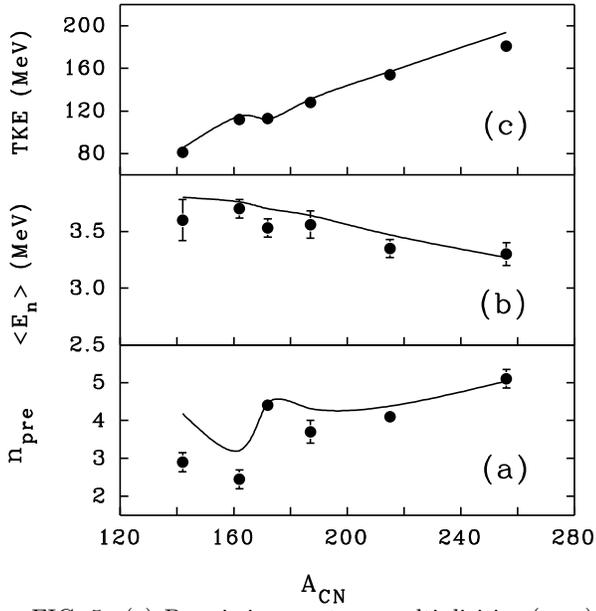}

\caption{(a)  Prescission  neutron  multiplicities  ($n_{pre}$),  (b)  mean
energy of the evaporated neutrons ($<E_n>$) and (c) total kinetic energy of
the fragment (TKE), plotted as a function of A$_{CN}$. The solid curves are
the present calculations, and the filled circles are the corresponding data
\protect\cite{e3}.}

\label{fig_ke}
\end{figure}

 To  have a closer look into the predictions of the present model so far as
other  related  observables  are  concerned,  the  average  energy  of  the
prescission  neutrons,  $<E_n>$  and  total  kinetic  energy  (TKE)  of the
fragments have been plotted as a function of  the  compound  nuclear  mass,
A$_{CN}$, in Figs.~\ref{fig_ke}(b) and \ref{fig_ke}(c), respectively. It is
seen  from  the  figure  that  the  theoretical  predictions  for  all  the
observables ({\it solid curve}) are in good agreement with  the  respective
experimental  data  ({\it  filled  circles}). The observed system dependent
fluctuations of both $<E_n>$ and TKE are also, reproduced  fairly  well  in
the present calculations.

\subsection{Fragment mass asymmetry dependence}
\label{sec_asy}

\begin{figure}[h]
\centering
\epsfysize=8 cm
\epsffile[122 192 510 604]{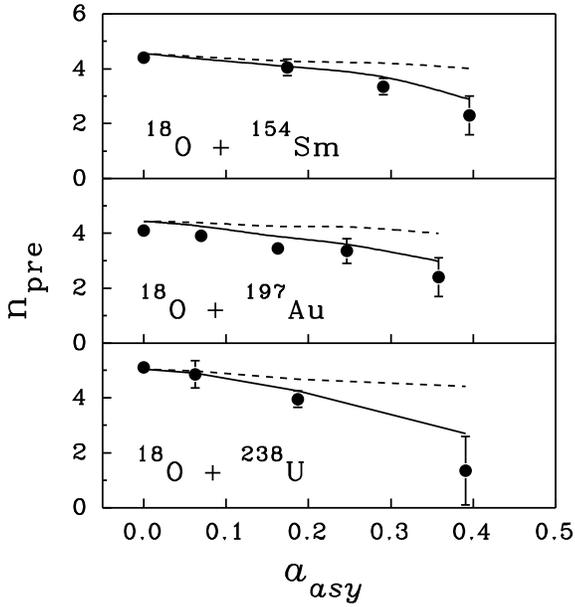}

\caption{Prescission  neutron  multiplicity  $n_{pre}$  as  a  function  of
fragment  mass  asymmetry  $a_{asy}  $for  $^{18}O$  induced  reactions  on
$^{154}Sm,  \  ^{197}Au  \ \rm{and} \ ^{238}U$. Filled circles correpond to
the experimental data \protect\cite{e3}.  The  dash  curves  represent  the
calculated  results  with  friction form factors of \protect\ref{eq.10} and
the solid curves represent the calculated results  with  modified  friction
form factors {\it(see text)}.}

\label{fig_asym}
\end{figure}

Calculations   have   also   been  performed  to  reproduce  the  available
experimental data for the asymmetric fission of  the  heavy  mass  compound
systems.  The  systems  studied  for  the  asymmetric  fission are the ones
produced in $^{18}O \  (E_{lab}  =  158.8  \  MeV)$  induced  reactions  on
$^{154}Sm$,  $^{197}Au$  and  $^{238}U$ \cite{e3}. Fig.\ref{fig_asym} shows
the predicted prescission neutron multiplicities $n_{pre}$ (dashed curves),
calculated with the same friction values as used in  symmetric  fission  in
the  previous  section,  as a function of the fragment asymmetry, alongwith
the corresponding experimental data (solid circles). It is evident from the
figure that, for all the systems, the predicted  values  of  $n_{pre}$  are
very  weakly  dependent on the fragment asymmetry, whereas the experimental
values decrease rapidly with the increase in fragment asymmetry. It may  be
conjectured   that   the   friction   form   factors  calculated  from  the
Werner-Wheeler prescription,  which  successfully  explains  the  symmetric
fission data, are rather inadequate in reproducing the experimental data in
the  case  of  asymmetric  fission. It appears that the above friction form
factors need some modification, ie, some extra asymmetry dependence  should
be  included  in the form factor to explain the asymmetric fission data. It
has been found that, with the inclusion of a factor $h(\alpha) =  \exp(-  K
\alpha^2)$  in  the expression for friction form factor (Eqn.~\ref{eq.10}),
the predicted values of $n_{pre}$ (solid curves) agree quite well with  the
respective experimental data. The value of the constant $K$ was found to be
$161\pm3$  which  is independent of the mass of the compound system. It is,
therefore, interesting to note that with the inclusion of this  extra  term
$h(\alpha)$,  we can still use the same value of the viscosity coefficient,
$\mu_0$, as used earlier in Secs.~\ref {sec_mul}, \ref{sec_tke} to  explain
the  prescission  neutron  multiplicity  data for both symmetric as well as
asymmetric fission.

\begin{figure}[h]
\centering
\epsfysize=8 cm
\epsffile[77 281 498 689]{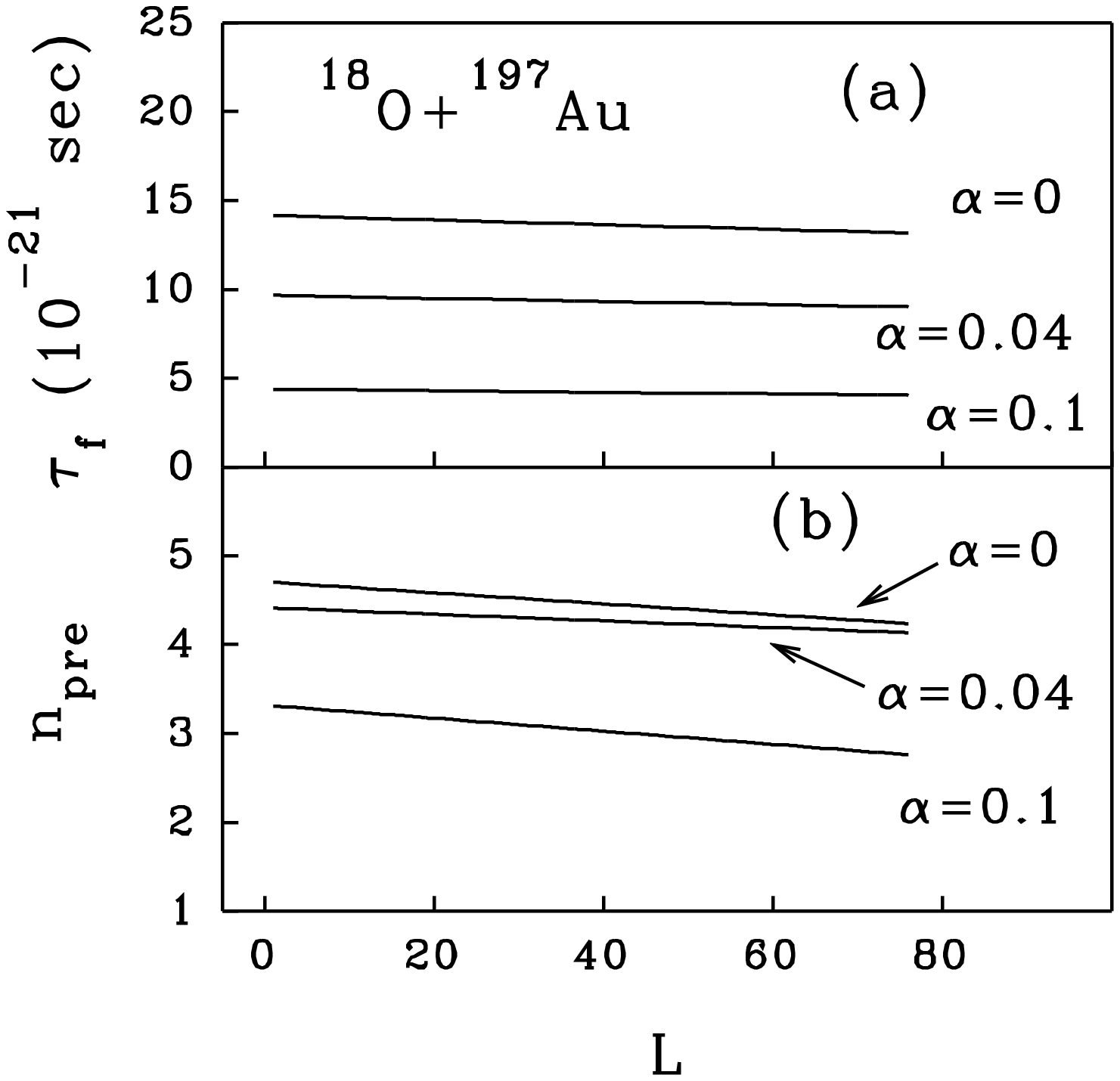}

\caption{  Theoretical  predictions  of  (a) Fission time $\tau_f$, and (b)
prescission neutron multiplicities  $n_{pre}$  plotted  as  a  function  of
angular  momentum $L$ for the reaction 158.8 MeV $^{18}O$ on $^{197}Au$ for
different values of the asymmetry parameter $\alpha$.}

\label{fig_lt}
\end{figure}

We have also studied the variation of prescission neutron multiplicities as
well  as the time scale of fission as function of the angular momentum. The
Figs. \ref{fig_lt}(a, b) show  the  calculated  results  for  fission  time
$\tau_f$  and neutron multiplicities $n_{pre}$, respectively, for a typical
system $^{18}O$+$^{197}Au$ for  different  values  of  asymmetry  parameter
$\alpha$. It is clear from the figure that both $n_{pre}$ and $\tau_f$ have
only a weak dependence on $L$. However, both these quantities have a strong
dependence  on  fragment  asymmetry  and  are  found  to  decrease with the
increase in fragment asymmetry. The time scales vary in the range of 0.5 to
1.5 ($\times 10^{-20}$ sec.) as asymmetry parameter $\alpha$ decreases from
0.1 to 0. A similar trend is observed in the variation of $n_{pre}$.

\section{Summary and Conclusions}

\label{conclud}

To  sum  up,  we have developed a dynamical model for fission where fission
trajectories are generated by solving Euler-Lagrange  equations  of  motion
using  a combination of one- and two-body frictions. Evolution of shapes of
the fissioning  nuclei  have  been  computed  from  the  generalised  shape
parametrisation  of Brack {\it et. al.} \cite{bra}. Different friction form
factors have been used for ground state to saddle and  saddle  to  scission
regions. In the ground state to saddle region, the one body 'wall' friction
has been used with an attenuation coefficient of 0.1 , whereas the two body
dissipative  forces,  derived  from  Werner-Wheeler prescription, have been
used in the saddle to scission region  with  the  value  of  the  two  body
viscosity  coefficient  $\mu_0  =  4  \times10^  {-23}MeV  \cdot  sec \cdot
fm^{-3}$. Same values of the friction coefficients have been used  for  all
the  systems  considered  here. With these values of friction coefficients,
the  typical  time  scales  of  fission  as  obtained  from   the   present
calculations  were  $\sim  \  (1-2)  \times  10^{-20}$  sec.  for symmetric
fission,  which  are  similar  to  the  values  reported  earlier  in   the
literature.  Emission  of  neutrons along the fission trajectories has been
simulated through Monte-Carlo simulation technique. The  evolution  of  the
fission  trajectories  has  been corrected for prescission proton emission,
which has been simulated in a similar  way  as  it  was  done  for  neutron
emission.  However,  the  prescission proton emission was found to be quite
small compared to the neutron emission for the systems  considered  in  the
present  studies.  The  emission  of complex particles ( d, t, $^4$He
etc.) and their effects on prescision neutron emission has  been  neglected
in the present calculation as it is difficult to incorporate them in the
present model because of their composite nature. Generally complex particle
emission is much less
as compared to the neutron emission; however, for some fissioning nuclei
$\alpha$ emission may be relatively favoured due to structure effects and it
may affect the neutron multiplicity and other fission observables.
 Theoretical predictions for prescission neutron
multiplicities,  total  kinetic  energies  and  the  mean  energies  of the
prescission neutrons, for all the systems considered here, agree quite well
with the corresponding available experimental data.

In  the  present  model,  the  fragment  mass  asymmetry  dependence of the
prescission neutron multiplicities and other physically relevent quantities
have also been studied. It has been found that with  the  inclusion  of  an
extra  factor  ($\exp(-K  \alpha^2)$,  $K$=161$\pm$3)  in the friction form
factor (Eqn.~\ref{eq.10}), the observed decrease of the prescission neutron
multiplicities with the increase in fragment mass asymmetry, are  explained
fairly  well  with  the  same value of the viscosity coefficient $\mu_0$ as
used in the case of symmetric fission. Further, it has been  observed  that
the  prescission  neutron multiplicities as well as fission time are weakly
dependent on the angular momentum of the fissioning  system.  However,  the
fission  times  are  found  to be strongly dependent upon the fragment mass
asymmetry and decrease from 1.5 to 0.5 ($\times 10^{-20}$ sec.) as the  the
asymmetry parameter $\alpha$ increase from 0.0 to 0.1.

To   conclude,   in   the   present   model,  with  the  generalised  shape
parametrisation and modified  Werner-Wheeler  friction  form  factor,  both
symmetric  and  asymmetric splitting of the compound nucleus can be treated
on the same footing. The present model is quite  successful  in  explaining
prescission   neutron  multiplicities,  total  fragment  kinetic  energies,
average energies of the prescission neutrons, as  well  as  their  fragment
mass asymmetry dependence.

\pagebreak

\pagebreak

\end{document}